\documentclass{PoS}
\usepackage{epsfig}
\usepackage{amsmath}
\usepackage{amssymb}
\usepackage{amsbsy}
\usepackage{amsfonts}
\usepackage{graphics}
\usepackage{latexsym}
\usepackage{mathrsfs}
\usepackage{dcolumn}
\usepackage{wasysym}
\newcommand{\onesixth}{\mbox{\small $\frac{1}{6}$}}      % 1/6
\newcommand{\half}{\mbox{\small $\frac{1}{2}$}}          % 1/2
\newcommand{\bc}{\begin{center}}
\newcommand{\ec}{\end{center}}
\def\kreis{\raise0.85pt\hbox{$\scriptstyle\bigcirc$}}
\def\vollk{\lower0.85pt\hbox{\Large $\bullet$}}

\vspace*{-1.5cm}

\title{\hfill {\large \tt PoS(LATTICE 2013)499}\\[-0.5em]\hfill {\large \tt ADP-13-25/T845}\\[-0.5em]\hfill {\large \tt
  DESY 13-221}\\[-0.5em]\hfill{\large \tt Edinburgh 
    2013/33}\\[-0.5em]\hfill {\large \tt LTH 994}\\[2em] 
Electromagnetic splitting of quark and pseudoscalar meson masses from dynamical
QCD $\mathbf{+}$ QED} 
\ShortTitle{Dynamical 1+1+1 flavor QCD + QED}

\author{R. Horsley$^a$, Y. Nakamura$^b$, D. Pleiter$^c$,
  P.E.L. Rakow$^d$, \speaker{G. Schierholz}$^{e}$, H. St\"uben$^f$,
  R.D. Young$^g$ and
  J.M. Zanotti$^g$\\ \\ 
$^a$ School of Physics and Astronomy, University of Edinburgh, Edinburgh
EH9 3JZ, United Kingdom\\ 
$^b$ RIKEN Advanced Institute for Computational Science, Kobe, Hyogo
650-0047, Japan\\ 
$^c$ JSC, Forschungszentrum J\"ulich, 52425 J\"ulich, Germany\\ 
$^d$ Theoretical Physics Division, Department of Mathematical Sciences,
University of Liverpool, Liverpool L69 3BX, United Kingdom\\ 
$^e$ Deutsches Elektronen-Synchrotron DESY, 22603 Hamburg, Germany \\ 
$^f$ RRZ, University of Hamburg, 20146 Hamburg, Germany \\
$^g$ CSSM, School of Chemistry and Physics, University of Adelaide,
Adelaide SA 5005, Australia}
\author{QCDSF Collaboration}

\abstract{Lattice QCD simulations are now reaching a precision where
electromagnetic corrections from QED become important. In
investigating the effects of $\mathrm{SU(3)}$ breaking due  
to quark mass differences within QCD, a group-theoretical analysis of
the mass dependence greatly helped us organize our results. We now do
the same with electromagnetic charge effects by extending the
calculations to dynamical $1+1+1$ flavor QCD~+~QED.}

\FullConference{The XXXI International Symposium on Lattice Field Theory\\
                 July 29 - August 3, 2013\\
                 Mainz, Germany}

\begin{document}

\section{Introduction}

One of the most profound open questions in particle physics is to
understand the pattern of flavor symmetry breaking and mixing, and the
origin of CP violation. In~\cite{Bietenholz:2011qq} we have outlined a  
program to systematically investigate the pattern of flavor symmetry
breaking. The program has been successfully applied to meson and
baryon masses involving up ($u$), down ($d$) and strange ($s$) quarks. 
%In this project
%we wish to extend the investigations to include electromagnetic effects.

A distinctive feature of our simulations is the way we tune the light
and strange quark masses. We have our best theoretical understanding
when all three quark flavors have the same mass, because we can use
the full power of flavor $\mathrm{SU(3)}$. Starting from the
$\mathrm{SU(3)}$ symmetric point, our strategy is to keep the singlet
quark mass $\displaystyle \bar{m}=(m_u+m_d+m_s)/3$ fixed at its
physical value, while $\displaystyle \delta m_q \equiv m_q-\bar{m}, \,
q=u, d, s$ is varied. As we move from the symmetric point
$m_u=m_d=m_s$ (where the pion mass is $\sim 411\,\mbox{MeV}$) 
to the physical point along the path $\bar{m}=constant$,
the $s$ quark becomes heavier, while the $u$ and $d$ quarks become
lighter. These two effects tend to cancel in any flavor singlet
quantity. To leading order, the cancellation is exact at the symmetric
point, and we 
have found that it remains good down to the lightest points we have
simulated so far~\cite{Bietenholz:2011qq}.

%So far electromagnetic (EM) effects have widely been neglected. 
In order to compute physical observables to high precision, it is
important to include and control contributions from QED. Recent
lattice investigations of electromagnetic (EM) corrections to hadron
observables have been performed on pure QCD background
configurations~\cite{QED}, while a simulation with dynamical
photons, including meson-photon mixing effects, is still missing. In this
project we will extend our previous simulations of $2+1$ flavor QCD
with SLiNC fermions to a fully dynamical simulation of $1+1+1$ flavor
QCD~+~QED.  

\section{QCD~+~QED pseudoscalar meson mass formulae}

In pure QCD~\cite{Bietenholz:2011qq} our strategy was
to start from a point with all three sea quark 
masses equal, $m_u=m_d=m_s$, and extrapolate towards the physical
point by keeping the average sea quark mass $\bar{m} =(m_u+m_d+m_s)/3$
constant. For this trajectory to reach the physical point we 
start at a point $\bar{m}=m_0$, where $M_\pi=M_K$ with
$2M_K^2+M_\pi^2$ equal to its physical value. That is
$M_\pi=M_K=413\,\mbox{MeV}$. We call this point the physical SU(3)
symmetric point. We denote the distance from $m_0$ by $\delta
m_q=m_q-m_0$, $q=u,d,s$. This forms a plane, as we have the constraint
$\delta m_u+\delta m_d+\delta m_s =0$. The bare quark masses are
defined by 
\begin{equation}
%\begin{split}
am_0 = \frac{1}{2\kappa_0}-\frac{1}{2\kappa_c} \,,\quad
am_q = \frac{1}{2\kappa_q}-\frac{1}{2\kappa_c} \,,
%\end{split}
\end{equation}
where $\kappa_0$ gives the quark mass at the physical SU(3) symmetric point,
and where vanishing of all quark masses along the line
$\kappa_u=\kappa_d=\kappa_s$ determines $\kappa_c$. The quark masses $m_q$
are subject to additive and multiplicative renormalization, while the
reference point $m_0$ gets multiplicatively renormalized
only~\cite{Bietenholz:2011qq}. 

In this presentation we shall concentrate on the pseudoscalar meson octet. The
expansion around $m_q=m_0$, valid for the outer ring of the
pseudoscalar octet, was found to be~\cite{Bietenholz:2011qq}
\begin{equation}
\begin{split}
M^2(a\bar{b})&=M_0^2+ \alpha(\delta m_a +\delta m_b)\\ &+ \beta_0\onesixth(\delta m_u^2 +
  \delta m_d^2 + \delta m_s^2) + \beta_1(\delta m_a^2 + \delta m_b^2) +
  \beta_2(\delta m_a -\delta m_b)^2 
\end{split}
\label{masses}\end{equation}
for arbitrary quarks $q=a,b$, with $\alpha$ and $\beta_0, \beta_1,
\beta_2$ being the LO and NLO expansion coefficients, respectively.

It is useful, in many respects, to vary valence and sea quark masses
independently. This is referred to as partial quenching (PQ). In this
case the sea quark masses remain constrained by $\bar{m}=constant$,
while the valence quark masses $\mu_u, \mu_d, \mu_s$ are
unconstrained. Defining $\delta\mu_q=\mu_q-\bar{m}$, we obtain the
PQ mass formula
\begin{equation}
\begin{split}
M^2(a\bar{b})&=M_0^2+ \alpha(\delta \mu_a +\delta \mu_b)\\ &+
\beta_0\onesixth(\delta m_u^2 + \delta m_d^2 + \delta m_s^2) + \beta_1(\delta
\mu_a^2 + \delta \mu_b^2) + \beta_2(\delta \mu_a -\delta \mu_b)^2 
\end{split}
\label{pqmasses}\end{equation}
When $\mu_q \rightarrow m_q$, this result reduces to the previous
result (\ref{masses}). The coefficients that appear in the expansion
about the flavor symmetric point (\ref{masses}) and in the PQ case
(\ref{pqmasses}) are the same. Hence this offers a computationally cheaper
way of obtaining them.

\begin{figure}[b]
   \begin{center} 
      \includegraphics[width=4.00cm,angle=90,clip]{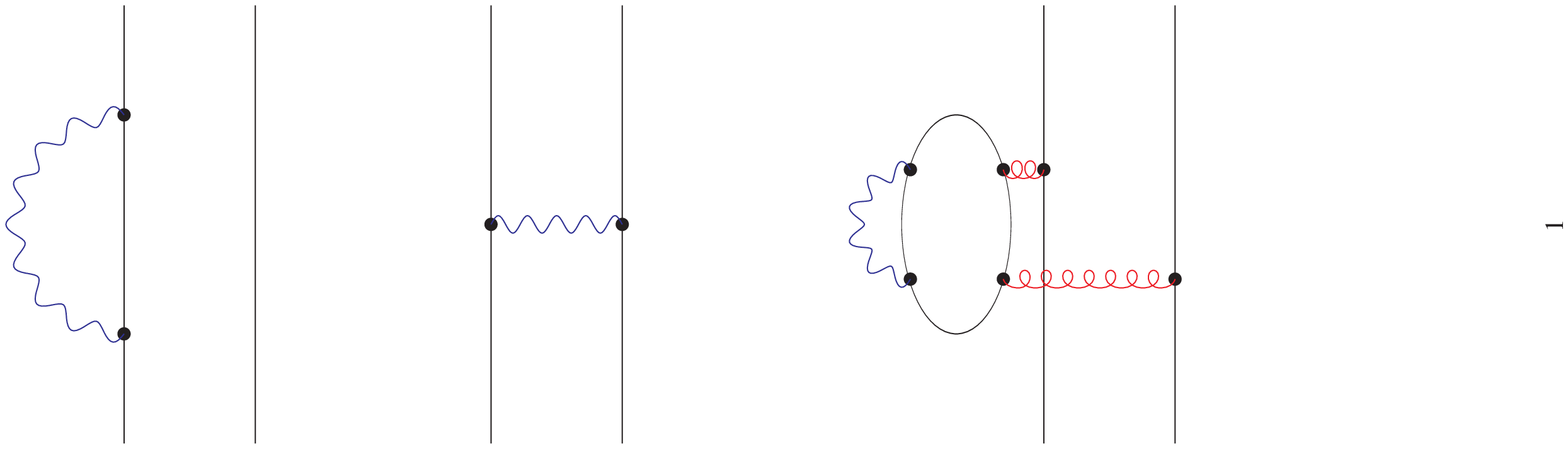}
   \end{center} 
\caption{Examples of Feynman diagrams contributing to the 
         meson electromagnetic mass to order $e^2$.}
\label{diag1}
\end{figure}

The symmetry of the electromagnetic current is similar to the symmetry
of the quark mass matrix. The simplifications that come from the
constraint $\delta m_u +\delta m_d +\delta m_s = 0$ in the mass case
are similar to the simplifications we get from the 
identity $e_u + e_d + e_s = 0$. One difference between quark mass
and electromagnetic expansions is that in the mass
expansion we can have both odd and even powers of $\delta m_q$,
whereas we are only allowed even powers of the quark charges. We can
therefore read off the leading QED corrections
from~\cite{Bietenholz:2011qq}, dropping the linear terms and
changing masses to charges. For the outer mesons, and also for the
partially quenched $q\bar{q}$ mesons with all annihilation diagrams
turned off, we find
\begin{equation}
\begin{split}
M^2(a\bar{b})&=M_0^2+ \alpha(\delta \mu_a +\delta \mu_b)\\ &+
\beta_0\onesixth(\delta m_u^2 + \delta m_d^2 + \delta m_s^2) + \beta_1(\delta
\mu_a^2 + \delta \mu_b^2) + \beta_2(\delta \mu_a -\delta \mu_b)^2
\\ &+ \beta_0^{\rm EM}(e_u^2+e_d^2+e_s^2) + \beta_1^{\rm
  EM}(e_a^2+e_b^2) + \beta_2^{\rm EM}(e_a-e_b)^2 \\ &+
\gamma_0^{\rm EM}(e_u^2\delta m_u+e_d^2\delta m_d+e_s^2\delta m_s)
+\gamma_1^{\rm EM}(e_a^2\delta \mu_a+e_b^2\delta \mu_b) \\ &+
\gamma_2^{\rm EM} (e_a-e_b)^2(\delta \mu_a+\delta \mu_b) +
\gamma_3^{\rm EM} (e_a^2-e_b^2)(\delta \mu_a-\delta \mu_b) \,. 
\end{split}
\label{mixedmasses}\end{equation}
The coefficients in (\ref{mixedmasses}) can be matched up with
different classes of Feynman diagrams shown in Fig.~\ref{diag1}. The
first diagram, with both ends of the photon attached to the
same valence quark, contributes to $(\beta_1^{\rm EM}+\beta_2^{\rm
  EM})$ as well as $(\gamma_1^{\rm EM}+\gamma_2^{\rm EM}+\gamma_3^{\rm
  EM})$. The second diagram, with the photon crossing between the
valence lines, only 
contributes to $\beta_2^{\rm EM}$ and $\gamma_2^{\rm EM}$. The last
diagram, with the photon 
being attached to the sea quarks, is an example of a diagram
contributing to $\beta_0^{\rm EM}$ and $\gamma_0^{\rm EM}$. It would
be missed out if the 
electromagnetic field was quenched instead of dynamical. 

Except for $\beta_0$, $\beta_0^{\rm EM}$ and $\gamma_0^{\rm EM}$, all
coefficients can be determined by PQ simulations at our expansion
point. The term $\displaystyle \beta_0^{\rm EM}(e_u^2+e_d^2+e_s^2)$
can be absorbed into $M_0^2$. The coefficients $\beta_0$ and
$\gamma_0^{\rm EM}$ require simulations with unequal sea quark masses.
Many of the terms in (\ref{mixedmasses}) cancel in the combination
\begin{equation}
\begin{split}
&M^2(a\bar{b})-\left[M^2(a\bar{a})+M^2(b\bar{b})\right]/2 =
\beta_2(\delta \mu_a-\delta \mu_b)^2 
+ \beta_2^{\rm EM}(e_a-e_b)^2\\ &\hspace*{2.94cm}+\gamma_2^{\rm EM} (e_a-e_b)^2(\delta
\mu_a+\delta \mu_b) + \gamma_3^{\rm EM} (e_a^2-e_b^2)(\delta
\mu_a-\delta \mu_b) 
\end{split}
\end{equation}
that will be important in our later discussions.

\section{Lattice setup}

\begin{figure}[b]
   \begin{center} \vspace*{0.25cm}
      \epsfig{file=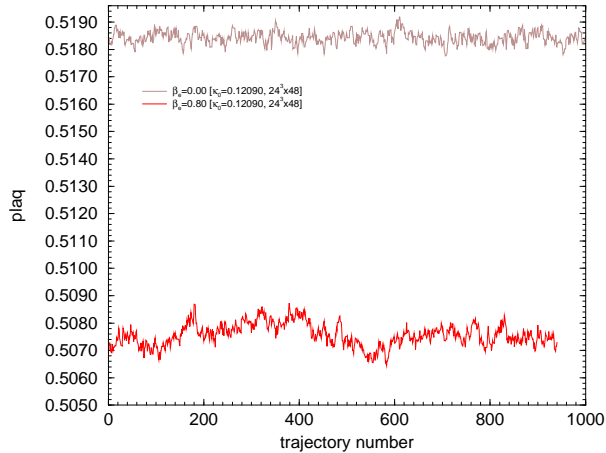,width=8cm}
   \end{center}  
\caption{The average plaquette for $\beta=5.50$ and
  $\kappa_u=\kappa_d=\kappa_s=0.12090$ on the $24^3\times 48$ lattice
  for $e^2=1.25$ (bottom red line) and $e^2=0$ (top gray line) as a
  function of trajectory number.}
\label{plaq}
\end{figure} 

The action we are using is  
\begin{equation} 
   S = S_G + S_A + S_{F}^u + S_{F}^d + S_{F}^s \,.
\label{action}
\end{equation}
Here $S_G$ is the tree-level Symanzik improved SU(3) gauge action,
and $S_A$ is the noncompact U(1) gauge action~\cite{Gockeler:1989wj}
of the photon,
\begin{equation}
S_A=\frac{1}{2e^2} \,\sum_{x, \mu<\nu} \left(A_\mu(x) + A_\nu(x+\mu)
- A_\mu(x+\nu) - A_\nu(x)\right)^2 \,.
\end{equation}
The fermion action for each flavor is
\begin{eqnarray}
   \tilde{S}_{F}^q &=& \sum_x \Big\{{1 \over 2} 
                        \sum_{\mu}\big[\overline{q}(x)(\gamma_\mu - 1)
                          e^{-ie_q\,A_\mu(x)}\tilde{U}_\mu(x) q(x+\hat{\mu}) -
                        \overline{q}(x)(\gamma_\mu + 1) 
                          e^{ie_q\,A_\mu(x)}\tilde{U}^\dagger_\mu(x-\hat{\mu})
                          q(x-\hat{\mu})\big] 
                        \nonumber   \\
            && \hspace*{5.01cm}+ {1 \over 2\kappa_q} \overline{q}(x)q(x) -
             {1 \over 4} c_{SW} \sum_{\mu\nu}
                 \overline{q}(x)\sigma_{\mu\nu}F_{\mu\nu}(x)q(x)
                                        \Big\} \,,
\label{faction}
\end{eqnarray}
where $\displaystyle \tilde{U}_\mu$ is a single iterated mild stout
smeared link~\cite{Bietenholz:2011qq}. The clover coefficient $c_{SW}$
has been computed nonperturbatively~\cite{Cundy:2009yy}. The quark
charges are $e_u=2/3$ and $e_d=e_s=-1/3$ (in units of $e$). We
presently neglect EM modifications to the clover term. This will leave
us with corrections of $O(e^2 a)$, which are presumably smaller than
the $O(a^2)$ corrections from QCD. 

Upon integrating out the Grassmann variables in the partition
function, and rewriting the resultant determinant using
pseudofermions, the effective action reads (generically)
\begin{equation}
\begin{split}
   S[U,A,\{\phi^\dagger,\phi\}]
      &= S_G[U] + S_A[A] +
        \phi_u^\dagger \left[ 
                        {\cal M}(\kappa_u)^\dagger\!\! {\cal M}(\kappa_u)
                     \right]^{-\half}\phi_u \\ &+
        \phi_d^\dagger \left[ 
                        {\cal M}(\kappa_d)^\dagger\!\! {\cal M}(\kappa_d)
                     \right]^{-\half}\phi_d +
        \phi_s^\dagger \left[
                        {\cal M}(\kappa_s)^\dagger\!\! {\cal M}(\kappa_s) 
                    \right]^{-\half}\phi_s \,,
\end{split}\end{equation}
where ${\cal M}$ is the fermion matrix. We deal with the square root
of ${\cal M}^\dagger\!\! {\cal M}$ by rewriting it as a rational function
\begin{equation}
X^{- 1/n} = \alpha_0 + \sum^N_{k=1} \frac{\alpha_k}{X + \beta_k} 
\label{HMC}
\end{equation}
and employ the Rational Hybrid Monte Carlo (RHMC)
algorithm~\cite{Clark:2006fx}. At the 
symmetric point, $\kappa_u=\kappa_d=\kappa_s$, this reduces to $2+1$
quark species. Then the Hybrid Monte Carlo (HMC) algorithm can be used
for the $d$ and $s$ 
quarks, while the RHMC algorithm is used for the $u$ quark. Away from
the symmetric point we would not expect to run into a sign problem as
we will always keep $m_u \approx m_d$.  

\section{Preliminary results and discussion}

Our first dynamical QCD~+~QED simulation was done on the $24^3\times
48$ lattice at $\beta=5.50$ and $\kappa_u=\kappa_d=\kappa_s =
0.12090$. That is at the flavor symmetric point $\delta m_u=\delta
m_d=\delta m_s=0$. We chose $e^2 = 1.25$. In Fig.~\ref{plaq} we
compare the average plaquette with and without dynamical photons. The
difference is significant. Our strategy is to simulate 
at an artificially large coupling $\alpha_{\rm EM} \approx 1/10$, and
then interpolate between this point and pure QCD to the physical
coupling $\alpha_{\rm EM} = 1/137$. 

\begin{figure}[t]
   \begin{center} \vspace*{-1.0cm}
      \epsfig{file=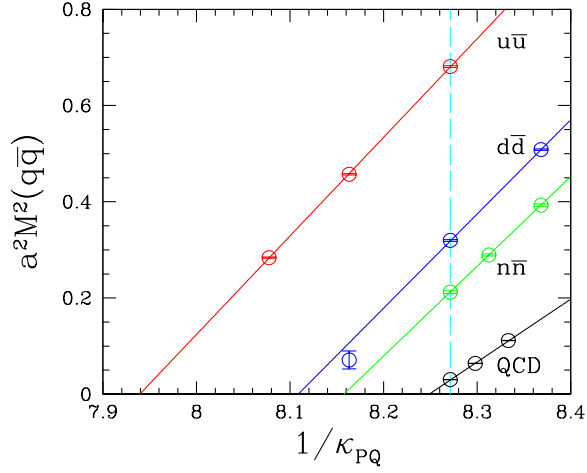,width=8cm}
   \end{center}  
\caption{Partially quenched QCD~+~QED pseudoscalar meson masses $aM(q\bar{q})$
  for $q\bar{q}=u\bar{u}$, $d\bar{d}$ and $n\bar{n}$ against
  $1/\kappa_{\rm PQ}$. Also shown are the PQ masses from pure QCD, as
  given in~\cite{Horsley:2011wr}.} 
\label{pqfig}
\end{figure}

As a first application we have looked at the EM mass shifts of quark
and pseudoscalar meson masses. In Fig.~\ref{pqfig} we show PQ masses
$aM(q\bar{q})$ for $q\bar{q}=u\bar{u}$, $d\bar{d}$ $(=s\bar{s})$ and a
fictitious electrically neutral quark $n$, $q\bar{q}=n\bar{n}$, as a
function of the PQ hopping parameter $\kappa_{\rm PQ}$. 
The first point to notice is that the mesons have become much heavier,
especially the $u\bar{u}$. We attribute this mostly to a shift in $\kappa_c$
for the quarks, due to their electromagnetic self-interaction, which
amounts to an additive quark mass renormalization. We
would obviously expect this to be a bigger effect for the $u$ than for 
the $d$ or $s$ quark, as observed. At the flavor symmetric point,
$\kappa_{\rm PQ} = 0.1209$, we find
\begin{equation}
\begin{split}
aM(n\bar{n})=0.4606(30)\,&,\quad aM(d\bar{d})=0.5655(16)\,,\\
aM(u\bar{d})=0.7310(15)\,&,\quad aM(u\bar{u})=0.8283(11) \,.
\end{split}
\end{equation}
This is to be compared with the corresponding mass of pure QCD,
$aM=0.1779(6)$~\cite{Bietenholz:2011qq}. We estimate the
lattice spacing $a$ to be $\approx 10\%$ smaller than in pure QCD,
using the vector meson mass for determining the change in scale.

A reasonable definition of the additive quark mass renormalization for
each flavor is 
\begin{equation}
\Delta am_q=
\frac{1}{2\kappa_c^{\phantom{q}}}-\frac{1}{2\kappa_{c\;{\rm PQ}}^q}
\,,  
\end{equation}
where $\kappa_c=0.121252$ is the critical hopping parameter of QCD, and
the PQ critical hopping parameter $\kappa_{c\;{\rm PQ}}^q$ can be read off
from Fig.~\ref{pqfig}. We find
\begin{equation}
%\begin{split}
\Delta am_n = 0.036\,e^2\,,\quad
\Delta am_d = \Delta am_s =0.056\,e^2\,,\quad 
\Delta am_u = 0.122\,e^2\,.
%\end{split}
\end{equation}
This is to be compared with the quark mass of pure QCD at the flavor
symmetric point, $am_u=am_d=am_s=0.012$. Note that $(am_u-am_n) :
(am_d-am_n) \approx 4 : 1$, as expected. 

Our present fits give $a^2\beta_0^{\rm EM}=1.20\,e^2$
and $a^2\beta_1^{\rm EM}=0.44\,e^2$, assuming a linear dependence on
$e^2$. Both $\beta_0^{\rm EM}$ and $\beta_1^{\rm EM}$ come almost
entirely from the shifts in $\kappa_c$. From PCAC and the leading
flavor expansion we expect  
that $M^2(u\bar{d})-\left[M^2(u\bar{u})+M^2(d\bar{d}\right]/2=0$. Violations
of this relation cannot be present at leading order in the quark
mass. In our data the $\beta_0^{\rm EM}$ and $\beta_1^{\rm EM}$ terms
cancel, and the only term which contributes at the expansion point
$\delta \mu_u=\delta \mu_d =\delta \mu_s=0$ is $\beta_2^{\rm EM}$,
\begin{equation}
M^2(u\bar{d})-\left[M^2(u\bar{u})+M^2(d\bar{d}\right]/2=\beta_2^{\rm
  EM}\,,
\end{equation}
corresponding to the middle diagram in Fig.~\ref{diag1}. From
our fits we obtain $a^2\beta_2^{\rm EM}= +0.025\,e^2$. 

In order to understand the sign of $\beta_2^{\rm EM}$, we note that
opposite charges attract and like charges repel. As a result, we would
expect EM effects to raise the mass of, for example, the $u\bar{d}$
($\pi^+$) meson relative to the $u\bar{u}$ and $d\bar{d}$
mesons. That is exactly what we find, which is mirrored in a positive
sign of $\beta_2^{\rm EM}$. We should, however, be aware that this
result might be contaminated by QCD and heavy quark effects.  

Besides the mass splittings of mesons and baryons, we are interested
in the masses of $u$, $d$ and $s$ quarks.  A point to make is that the
renormalization factors will now depend on both the QCD and the QED
coupling, and the $u$ quark will have a different renormalization
factor and anomalous dimension from the other two quarks. This means
that the ratio $m_d/m_u$ now depends on renormalization scheme and
scale. Likewise, isospin-violating mass splittings, such as $M_n-M_p$,
are scheme 
independent, but the question of how much of the 
splitting is due to the quark mass differences, and how much is due to
EM effects, becomes dependent on scheme and scale.

\section{Outlook}

In pure QCD we can impose perfect SU(3) symmetry by making all three
$\kappa$ values equal. With QED present, there is no way to have
perfect SU(3) symmetry with physical charge ratios. A physically
reasonable definition is to look 
for a line, where the neutral pseudoscalar masses $M(s\bar{d})$,
$M(d\bar{s})$ as well as the PQ flavor diagonal masses $M(u\bar{u})$, 
$M(d\bar{d})$, $M(s\bar{s})$ (with annihilation diagrams turned
off) and $M(n\bar{n})$ are equal. We are currently using PQ
calculations to locate this line. The line will have 
$\kappa_d=\kappa_s \neq \kappa_u$. 

This symmetric line will end at a point, where all neutral
pseudoscalar mesons are massless. We define this to be the chiral
point. It is the point, where all our quark masses are zero. In the
case of the $d$ and $u$ quarks this is the correct definition. Even
with QED present, we have a chiral SU(2) symmetry connecting $d$ and
$s$ quarks. So, if both quarks are massless, there will be a massless
Goldstone boson from the spontaneous symmetry breaking. Although the
neutral pseudoscalar mesons will be massless at the chiral point, the
charged mesons can have a mass from EM effects. Furthermore, the
charged axial vector currents are no longer conserved after QED is
added to the action. Hence, there is no Goldstone boson for the
charged pseudoscalar sector.  

To summarize, our strategy is to compute hadron observables, both in
QCD (which we have done already) and in QCD~+~QED with $\bar{m}$, the
average sea quark mass, to be the same (or nearly the same) in both
simulations. This we achieve by simulating at points, where the
QCD~+~QED pseudoscalar mesons have (approximately) the same mass as in 
the pure QCD simulation. To obtain statistically significant results,
the calculations are performed at a suitable value of $e^2$. We then
may interpolate the numbers to $\alpha_{\rm EM}=1/137$, knowing the
results at $e^2=0$. 

\section*{Acknowledgement}

This work has been supported in part by the EU under contract 227431
(HadronPhysics2) and contract 283286 (HadronPhysics3), and the
Australian Research Council by grants FT120100821 (RDY) and
FT100100005 (JMZ). The numerical simulations have been performed on 
JUQUEEN at JSC (J\"ulich) and DIRAC 2 at EPCC (Edinbugh), as well as on
ICE at HLRN (Berlin and Hannover). 
%We thank all funding agencies for support.  

\end{document}